\documentclass[12pt]{iopart}
\begin{document}

\title[L\'evy-Leblond and Schr\"odinger equations]
{Fundamental dynamical equations for spinor wave functions.
I. L\'evy-Leblond and Schr\"odinger equations}

\author{R. Huegele, Z.E. Musielak and J.L. Fry} 

\address{Department of Physics, The University of Texas 
at Arlington, Arlington, TX 76019, USA}

\ead{randy.huegele@mavs.uta.edu; zmusielak@uta.edu; fry@uta.edu}

\begin{abstract}
A search for fundamental (Galilean invariant) dynamical equations for 
two and four-component spinor wave functions is conducted in Galilean 
space-time.  A dynamical equation is considered as fundamental if it 
is invariant under the symmetry operators of the group of the Galilei 
metric and if its state functions transform like the irreducible 
representations of the group of the metric.  It is shown that there are 
no Galilean invariant equations for two-component spinor wave functions.  
A method to derive the L\'evy-Leblond equation for a four-component spinor 
wave function is presented.  It is formally proved that the L\'evy-Leblond 
and Schr\"odinger equations are the only Galilean invariant four-component 
spinor equations that can be obtained with the Schr\"odinger phase factor.  
Physical implications of the obtained results and their relationships to 
the Pauli-Schr\"odinger equation are discussed.  
\end{abstract}

%
\maketitle

\section{Introduction}

A physical theory of free particles is considered fundamental if 
its dynamical equations have the same form in all isometric frames 
of reference.  All coordinate transformations that do not change 
a given metric of space-time form a representation of the group 
of the metric.  In order for two observers with the same metric 
to identify the same particle, a state function describing this 
particle must transform like one of the irreducible representations 
(irreps) of the group of the metric.  

This definition was first formally introduced by Wigner$^1$, who 
determined all unitary irreps of the Poincar\'e group$^2$ and used 
them to classify the elementary particles in Minkowski space-time.  
The original Wigner work was used by Bargmann and Wigner$^3$ to 
obtain Poincar\'e invariant dynamical equations associated with 
each representation.  As a result, the Klein-Gordon$^{4,5}$, 
Dirac$^6$ and Proca$^7$ equations were formally obtained.  Local 
symmetries of these equations were discussed by Fushchich and 
Nikitin$^8$.
    
The situation is more complicated in Galilean space-time because 
vector irreps of the Galilei group of the metric have no physical 
interpretation$^9$.  However, there is an infinite number of 
projective (ray) irreps, which are different from the vector 
irreps of the group$^{10}$.  Typically, these projective irreps 
are determined by the method of induced representations$^{11,12}$, 
and they are characterized by a constant that enters a phase 
factor in defining the projective irreps$^2$.  The process of 
introducing the constant is the central extension of the Lie 
algebra and the corresponding group is called the extended 
Galilei group$^{2,8}$.  

This group had played an important role in the L\'evy-Leblond 
work$^{13,14}$, who considered free particles with arbitrary spin 
in Galilean space-time and represented these particles by a 
four-component spinor state function.  He used the Bargmann-Wigner 
method to derive a dynamical equation that describes evolution of 
this function in time and space, and demonstrated that the derived 
equation was Galilean invariant.  The obtained equation is now known 
as the L\'evy-Leblond equation$^8$.  Moreover, L\'evy-Leblond$^{14}$ 
also derived the Pauli-Schr\"odinger (PS) equation$^{15,16}$ by adding 
the electromagnetic field to his Galilean invariant equation.  The 
fields were made Galilean invariant by dropping the Maxwell term from 
Maxwell's equations.  He obtained the PS equation for a two-component 
spinor wave function, however, he did not investigate Galilean 
invariance of this equation.

Extensive studies of the L\'evy-Leblond and Pauli-Schr\"odinger equations 
were performed by Fushchich and Nikitin$^8$, who derived those equations
and discussed their local symmetries.  The authors also 'deduced' the 
Schr\"odinger equation$^{17}$ by assuming a form of their partial 
differential operator that directly leads to this equation, and showed 
that the equation was Galilean invariant; however, see also Refs (18-20).  
Actually, they demonstrated that the Schr\"odinger equation was the only 
Galilean invariant equation for a scalar wave function.  

In our work on free and spinless particles described by scalar and analytic 
state functions in Galilean space-time, we used the Principle of Galilean 
Relativity and the Principle of Analyticity to formally derive Schr\"odinger-like equations$^{21}$.  The Principle of Relativity requires that all inertial observers 
formulate the same physical law and that they identify the same elementary particle.  
According to this Principle, all dynamical equations describing evolution of the 
wave function in time and space must be invariant with respect to all transformations 
that leave the metric unchanged.  The Principle of Analyticity demands that the wave 
function describing elementary particles is analytic.  This shows that our method 
greatly differs from that used by Fushchich and Nikitin$^8$.  Similarly, our formal 
proof of the uniqueness of the Schr\"odinger equation$^{22}$ is also significantly 
different from that presented by Fushchich and Nikitin.  In our further development, 
we demontrated how to formulate fundamental (Galilean invariant) theories of waves 
and particles without using the concept of classical mass$^{23}$.  

More recently, de Montigny, Niederle and Nikitin$^{24}$ considered Galilean invariant 
theories by constructing indecomposable finite-dimensional representations of the 
homogeneous Galilei group.  They used these representations to derive a general Pauli 
anomalous interaction term and deduce wave equations that describe interaction of 
Galilei particles with an external electric field.  Moreover, Niederle and Nikitin$^{25}$
derived Galilean invariant dynamical equations for massless and massive fields.  They 
showed that their Galilean invariant equations can also be obtained by contraction of 
known and new relativistic wave equations.

The main objective of this paper is to search for Galilei invariant dynamical 
equations for spinor wave functions.  The presented results are complementary to 
those obtained by Fushchich and Nikitin$^8$, de Montigny, Niederle and Nikitin$^{24}$, 
and Niederle and Nikitin$^{25}$, however, our approach is significantly different and 
it is also more fundamental in the physical sense.  The reasons are the following.  
First of all, we begin our search with a general partial differential equation whose 
form is justified by our previously obtained results$^{21-23}$, which are based on 
the Principle of Galilean Relativity and the Principle of Analyticity.  Second, we 
search for dynamical equations that are symmetric under the transformations of the 
extended Galilei group$^{2,8,13}$.  Third, our approach allows for other than 
Schr\"odinger's phase function$^{20}$.  Finally, we consider both two-component 
and four-component spinor wave functions.  Our motivation for considering the 
former is that such spinors are now widely used in General Relativity (GR) 
primarily through the work of Penrose$^{26,27}$.  
     
We define a dynamical equation to be fundamental if it has the following properties: 
(i) invariance under the symmetry operators of the group of the metric; (ii) no 
mixed time and space partial derivatives; and (iii) state functions that transform 
like the irreducible representations of the group of the metric.  All the results 
presented in this paper are obtained with Schr\"odinger's phase function$^{20}$.  
Other phase functions are also possible and they lead to new equations that will 
be presented and discussed in the next paper of this series. 
     
Among the results presented in this paper a special emphasis should be given to 
our formal proof that there are no Galilean invariant equations for two-component 
spinor wave functions.  Now, as long as four-component spinor wave functions are 
concerned, we derived the L\'evy-Leblond equation by using a different method than 
that originally used by L\'evy-Leblond$^{13,14}$ and by Fushchich and Nikitin$^8$,
and demonstrated that the L\'evy-Leblond equation is the only first-order Galilean 
invariant equation; our proof is significantly different than that of Fushchich 
and Nikitin.  We also proved that the Schr\"odinger equation is the only second-order 
fundamental dynamical equation in Galilean space-time, and that there are no other 
higher-order fundamental equations for two and four-component spinors.  Finally, 
we demostrated that the PS equation for two-component spinor wave functions is not 
fundamental and that this is a consequence of the fundamental Schr\"odinger equation
for four-component spinor state functions. 

The outline of the paper is as follows: the basic properties of Galilean relativity 
and some previously obtained results, including Schr\"odinger's phase function, are 
briefly described in Sec. II;  fundamental dynamical equations for two and four-component 
spinor wave functions are derived and discussed in Sec. III; relationship of the obtained 
results to the Pauli-Schr\"odinger equation is described in Sec. IV; and our conclusions 
are given in Sec. V.

\section{Galilean relativity and Schr\"odinger's phase function}

\subsection{Group of the Galilei metric}

Galilean space-time is defined by the Galilei metric: $ds_1^2 = dx^2+dy^2+dz^2$ 
and $ds_2^2=dt^2$, where $x$, $y$, $z$, and $t$ are spatial and time coordinates.  
The metric is invariant under a set of transformations that forms the Galilei group.  
The group may be decomposed into subgroups such that
\begin{equation}
G = \left[ T(1) \otimes R(3) \right] \otimes_s \left [ 
T(3) \otimes B(3) \right]\ ,
\label{IIeq1}
\end{equation}

\noindent
where $T(1)$, $R(3)$, $T(3)$, and $B(3)$ are the subgroups of translation in time, 
rotations in space, translations in space, and boosts respectively.  The direct 
product and semi-direct product are denoted $\otimes$ and $\otimes_s$.

The Galilean transformations can be used to relate the coordinate systems of 
two observers that are spatially rotated, translated, and boosted relative to 
one another.  A Galilean transformation can be defined by 
\begin{equation}
{{\vec x}\to {\vec x^{\ \prime}}} = R{\vec x}+{\vec v}t+{\vec a}
\hskip0.2in {\rm and} \hskip0.2in t \to t^{\prime} = t + b\ , 
\label{IIeq2}
\end{equation} 

\noindent
where $R$ is a rotation matrix, $\vec v$ is the velocity vector of a boost 
relating the two coordinate systems, and $\vec a$ is a spacial translation 
relating the two coordinate systems.  The inverse Galilean transformation is 
\begin{equation}
{\vec x}=R^{-1}{\vec x^{\ \prime}}-R^{-1}{\vec v}t'-R^{-1}\left({\vec  a}-{\vec v}b 
\right) \hskip0.2in {\rm and} \hskip0.2in t = t^{\prime} - b\ . 
\label{IIeq3}
\end{equation} 

\noindent
The chain rule can be used to determine how the differential operators transform 
under the Galilean transformation
\begin{equation}
{\frac{\partial }
{\partial t'}}=\frac{\partial t}{\partial t'}\frac{\partial }
{\partial t}+\frac{\partial x^i}{\partial t'}\frac{\partial }
{\partial x^i}=\frac{\partial }{\partial t}-R^{-1}_{ij}v_j \frac{\partial }
{\partial x^i} , 
\label{IIeq4}
\end{equation} 
and
\begin{equation}
{\frac{\partial }
{\partial x'^i}}=\frac{\partial t}{\partial {x'}^i}\frac{\partial }
{\partial t}+\frac{\partial x^j}{\partial {x'}^i}\frac{\partial }{\partial 
x^j}=R^{-1}_{ij}\frac{\partial }{\partial x^j}\ , 
\label{IIeq5}
\end{equation} 

\noindent
where $i = 1$, $2$ and $3$, and $j = 1$, $2$ and $3$.

It has been demonstrated for scalar wave functions that the Galilei group does 
not lead to any dynamical equations that satisfy the principles of analyticity 
and relativity$^{21}$.  The first principle requires that state functions are 
analytic and the second principle demands that dynamical equations governing 
the state function are Galilean invariant.  Therefore an additional symmetry $\left|\psi^*\psi\right|=\left|\psi'^*\psi'\right|$ must be added to the group 
of the metric.  

The expanded symmetry group called the extended Galilei group is the universal 
covering group of the Galilei group$^{11,12}$.  The extended Galilei group 
exhibits structure that is similar to the Poincare group$^{2,8,13,21}$.  The 
arguments used in Ref. (21) for scalar wave functions apply equally well to 
$n$-component state functions such as spinors and vectors.  Consequently, we 
begin this work with the extended Galilei group, which has the following 
structure 
\begin{equation}
G_e = [R(3) \otimes_s B(3)] \otimes_s [T(3+1) \otimes U(1)]\ , 
\label{IIeq6}
\end{equation} 

\noindent
where $U(1)$ is a one-parameter unitary group.  We consider only the 
proper isochronal subgroup $G_+^\uparrow$ of $G_e$ that omits the space 
and time inversions, which can be treated separately.
 
Measurements of the norm of the scalar state function must produce the 
same results for all observers related by the Galilei transformations.  
Hence, the resulting transformation of the wave function $\psi$ between 
two inertial frames of reference $(\vec x, t)$ and $(\vec x^{\prime}, 
t^{\prime})$ is 
\begin{equation}
\psi(\vec x,t) = e^{i\phi(\vec x^{\prime},t^{\prime})}\psi^{\prime}
(\vec x^{\prime}, t^{\prime})\ ,
\label{IIeq7}
\end{equation} 

\noindent
with $\phi(\vec x^{\prime},t^{\prime})$ being a phase function to be 
determined. 

\subsection{Schr\"odinger's phase function}

In the previous work, Musielak and Fry$^{21}$ used the Galilei group of 
the metric and the Principle of Analyticity and the Principle of Galilean 
Relativity to formally derive Schr\"odinger-like equations.  They concluded 
that the Galilei group was incomplete for forming a fundamental theory of 
free particles and that the necessary modifications of the group led to 
the extended Galilei group.  The derived Schr\"odinger-like equation can 
be written in the following form
\begin{equation}
i {\partial \psi \over \partial t} + {\omega \over k^2} \nabla^2 
\psi = 0\ ,
\label{IIeq8}
\end{equation}

\noindent
where $\omega$ and $k$ are the eigevalues of the translation operators in 
time and space.  Properties of the eigenvalue equations ensure that $\omega 
/ k^2 = 1/2M $ is a constant in all inertial frames of reference.  $M$ is 
refered to as the "wave mass" and is related to the classical mass through 
the Planck constant $m=\hbar M$ (see Refs 21-23).  

Galilean invariance of the Schr\"odinger-like equation under the transformations 
of the extended Galilei group requires a phase factor $e^{i \phi(\vec x^{\prime},
t^{\prime})}$ with the phase function given by  
\begin{equation}
\phi(\vec x^{\prime},t^{\prime}) = m \vec v \cdot \vec x^{\prime} + \frac{1}{2}
mv^2 t^{\prime}\ , 
\label{IIeq9}
\end{equation}

\noindent
where $\vec v$ is the constant velocity of one inertial frame of reference with
respect to the other.

Now, when the state function is a spinor it has two or more components and each 
component must satisfy the Schr\"odinger-like equation given by Eq. (\ref{IIeq8}), 
which means that the phase function given by Eq. (\ref{IIeq9}) must be used (see 
Refs 8,20).  We will call $\phi(\vec x^{\prime},t^{\prime})$ the Schr\"odinger 
phase function and our search for fundamental (Galilean invariant) equations for 
spinor state functions described in this paper will be exclusively based on this 
phase function.  Our approach allows us to consider other phase functions and 
these functions lead to new fundamental dynamical equations that will be 
presented and discussed in the second paper of this series.   

\section{Dynamical equations for spinor state functions}

\subsection{Method}

Let us consider the following first-order partial differential equation
\begin{equation}
\left[ B_1{{\partial} \over {\partial t}}+B_{2j}{{\partial} \over {\partial 
x_j}} + B_3 \right] \psi(\vec x,t)=0\ ,
\label{IIIeq1}
\end{equation}

\noindent
where $B_1$, $B_{2j}$, and $B_3$ are arbitrary $N\times N$ matrices that 
are assumed to be free of any dependence on the space and time coordinates, 
and $\psi$ is a $N$-component spinor state function.  The motivation for 
this general form of the above equation is given by our previously obtained 
results that were based on the Principle of Galilean Relativity and the 
Principle of Analyticity$^{21,22}$.  The results show that the first-order 
derivatives used in this equation are eigen-operators on the state function 
$\psi$, which transforms as an irreducible representation of the extended 
Galilei group, and that a term with a constant matrix times $\psi$ can 
always be added to the equation.    

Dynamical equations must be invariant under Galilean transformations, 
so we will require Galilean invariance of the first-order differential 
equation in order to derive a set of restrictions on the matrices $B$.  
Applying Galilean transformations to Eq. (\ref{IIIeq1}) and regrouping 
the terms, we obtain
\[
\left[B_1'{{\partial} \over {\partial t}}+(B_1'R_{jk}v_k+B_{2k}'R_{kj})
{{\partial} \over {\partial x_j}}\right]\psi(\vec x,t)
\]
\begin{equation}
\hskip1.0in +\left[-{{i}\over{2}}mv^2B_1'-imR_{jl}v_kR_{kl}B_{2j}'+ 
B_3'\right]\psi(\vec x ,t)=0\ .
\label{IIIeq2}
\end{equation}

\noindent
For Eq. (\ref{IIIeq1}) to be invariant under Galilei transformation, 
Eq. (\ref{IIIeq2}) must be of the same form.  This requirement leads 
to the following conditions on the set of matrices
\begin{equation}
B_1 = GB_1G^{-1}\ ,
\label{IIIeq3}
\end{equation}
\begin{equation}
B_{2j} =GB_1G^{-1}R_{ij}v_j+ GB_{2i}G^{-1}R_{ij}\ ,
\label{IIIeq4}
\end{equation}
and
\begin{equation}
B_3 = -{i \over 2}mv^2GB_1G^{-1}-imv_iGB_{2i}G^{-1}+ GB_3G^{-1}\ .
\label{IIIeq5}
\end{equation}

\noindent
The conditions will be now used to search for fundamental dynamical 
equations for two and four-component spinor wave functions.

\subsection{First-order equations for two-component spinors}

Our main result obtained for two-component spinor state functions
is given by the following proposition.

\bigskip
\noindent
{\bf Proposition 1.} {\it If a two-component spinor state function $\psi$ 
transforms as $\psi(\vec x,t) = e^{i\phi(\vec x^{\prime},t^{\prime})}
\psi^{\prime}(\vec x^{\prime}, t^{\prime})$, where $\phi$ is the 
Schr\"odinger phase function given by Eq. (\ref{IIeq9}), and $(\vec x, t)$ 
and $(\vec x^{\prime}, t^{\prime})$ are two inertial frames of reference, 
then there are no invariant dynamical equations describing evolution of 
this function in Galilean space-time.}

\bigskip
\noindent
{\bf Proof.} Applying rotations only (no boosts) in the conditions given by 
Eq. (\ref{IIIeq3}) through Eq. (\ref{IIIeq5}) constrains the matrices to the 
following forms: $B_1 = c_1 I$, $B_{2j} = c_2 \sigma_{j}$and $B_3 = c_3 I$, 
where $c_1$, $c_2$, and $c_3$ are arbitrary constants.  This demonstrates 
that there are first-order equations that are rotationally invariant.  To 
be Galilean invariant the equation must also be boost invariant.

It turns out that it is not possible to construct Galilei boost operators 
for two-component spinors.  In general, one may construct a boost matrix 
from the velocity parameters $v_j$ and boost generators $X_j$ as the 
exponential expression given by $B(v)=e^{i X_j v_j}$.  The generators of 
Galilei boosts must obey the following commutation relations of the Galilei 
group: $\left[ X_{\theta_i}, X_{\theta_j} \right]=i X_{\theta_k} \epsilon_{ijk}$, 
$\left[ X_{v_i}, X_{v_j} \right]= 0$ and $\left[ X_{v_i}, X_{\theta_j} \right] 
= i X_{v_k} \epsilon_{ijk}$. 

Because Galilei boosts commute, they form an Abelian subgroup and a one 
dimensional irreducible representation exists.  However, the composition 
of boosts and rotations is the result of a semi-direct product and requires 
boosts and rotations to obey the group composition law that can be written 
as
\[
G(a, b, v, R(\theta))=T(b)S(a)B(v)R(\theta)
\]
\[
=G(a_2, b_2, v_2, R_2)G(a_1, b_1, v_1, R_1)
\]
\begin{equation}
=G(R_2a_1+a_2-v_2b_1, b_1+b_2, R_2v_1+v_2, R_2R_1)\ ,
\label{IIIeq6}
\end{equation}

\noindent
which is composed of translations in time $T(b)$, translations in space $S(a)$, 
boosts $B(v)$, and rotations $R(\theta)$.  For the $2 \times 2$ generators of 
rotations $X_{\theta_i}=\sigma_i/2$, there are no $2 \times 2$ matrices that 
are able to satisfy the commutation relations as boost generators.

An interesting result is that this problem does not exit in the Minkowski 
space-time$^{26,28}$.  Hence, one may try to take the limit $c \rightarrow 
\infty$ of the Lorentz boost for two-component spinors.  Following Ref. 29, 
the result is  
\begin{equation}
B_{v_x}= \left( {\begin{array}{cc}
\cosh v_x/c & \sinh v_x/c \\
\sinh v_x/c & \cosh v_x/c \\
\end{array}} \right)\ ,
\label{IIIeq7}
\end{equation}
\begin{equation}
B_{v_y}=
\left( {\begin{array}{cc}
\cosh v_y/c & i\sinh v_y/c \\
-i\sinh v_y/c & \cosh v_y/c \\
\end{array}} \right)\ ,
\label{IIIeq8}
\end{equation}
and
\begin{equation}
B_{v_z}=
\left( {\begin{array}{cc}
e^{ v_z/c} & 0 \\
0 & e^{ v_z/c} \\
\end{array}} \right)\ ,
\label{IIIeq9}
\end{equation}

\noindent
which shows that the diverging matrix elements are obtained.  This is 
not surprising since the Galilei spinor boosts cannot be represented with 
$2 \times 2$ matrices.  Lorentz boosts do not commute as their Galilei 
counterparts.  As such the Lorentz group has a different universal covering 
group, SL(2,C) and it can be represented with $2 \times 2$ matrices$^{14}$.  
The physical implications of this result are that we cannot perform a 
Galilei boost of two component spinors.  This concludes the proof of 
Proposition 1.   

\subsection{First-order equations for four-component spinors}

After showing that there are no Galilean invariant dynamical 
equations for two-component spinor wave functions, we searched 
for fundamental dynamical equations describing evolution of 
four-component spinor wave functions in time and space.  The
obtained results are summarized by the following proposition.  

\bigskip
\noindent
{\bf Proposition 2.} {\it If a four-component spinor state function 
$\psi$ transforms as $\psi(\vec x,t) = e^{i\phi(\vec x^{\prime},
t^{\prime})} \psi^{\prime}(\vec x^{\prime}, t^{\prime})$, where 
$\phi$ is the Schr\"odinger phase function given by Eq. (\ref{IIeq9}),
and $(\vec x, t)$ and $(\vec x^{\prime}, t^{\prime})$ are two inertial 
frames of reference, then there is a Galilean invariant first-order 
partial differential equation for this function and the equation is 
known as the L\'evy-Leblond equation
\begin{equation}
\left[
\left( {\begin{array}{cc}
0 & 0 \\
I & 0 \\
\end{array}} \right)
 {{\partial} \over {\partial t}}+
\left( {\begin{array}{cc}
\sigma_j & 0 \\
0 & -\sigma_j \\
\end{array}} \right)
{{\partial} \over {\partial x_j}} +
\left( {\begin{array}{cc}
0 & 2imI \\
0 & 0 \\
\end{array}} \right)
\right] \psi(\vec x,t)=0
\label{IIIeq10}
\end{equation}

\noindent
where $j = 1$, $2$ and $3$, $\sigma_j$ are the $2 \times 2$ Pauli matrices, 
and $I$ is the $2 \times 2$ identity matrix.}

\bigskip
\noindent
{\bf Proof.} As in the case for two-component spinors, we seek a set of 
matrices that satisfy the conditions for invariance given by Eq. (\ref{IIIeq3})
through Eq. (\ref{IIIeq5}).  Let $B_1$ be an arbitrary $4 \times 4$ 
matrix, then
\begin{equation}
B_1 =
\left( {\begin{array}{cc}
P & Q \\
S & T \\
\end{array}} \right)\ ,
\label{IIIeq11}
\end{equation}

\noindent
where $P$, $Q$, $S$, and $T$ are arbitrary $2 \times 2$ matrices.  Applying an 
arbitrary rotation in the condition given by Eq. (\ref{IIIeq3}) results in 
four conditions on the $2 \times 2$ matrices $P$, $Q$, $S$, and $T$.  The
conditions are:  
\[
B_1 = RB_1R^{-1} = 
\left( {\begin{array}{cc}
U & 0 \\
0 & U \\
\end{array}} \right)
\left( {\begin{array}{cc}
P & Q \\
S & T \\
\end{array}} \right)
\left( {\begin{array}{cc}
U^{-1} & 0 \\
0 & U^{-1} \\
\end{array}} \right)
\]
\begin{equation}
=
\left( {\begin{array}{cc}
UPU^{-1} & UQU^{-1} \\
USU^{-1} & UTU^{-1} \\
\end{array}} \right)\ .
\label{IIIeq12}
\end{equation}

\noindent
Individually these four conditions are identical in form to 
that given by Eq. (\ref{IIIeq3}) for two-component spinors 
and the results are the same.  Therefore the $2\times 2$ 
matrices must be diagonal with arbitrary constant coefficients  
$p$, $q$, $s$, and $t$ and
\begin{equation}
B_1 =
\left( {\begin{array}{cc}
pI & qI \\
sI & tI \\
\end{array}} \right)\ .
\label{IIIeq13}
\end{equation}

For rotations only, the condition given by Eq. (\ref{IIIeq5}) has the 
same form as that given by Eq. (\ref{IIIeq3}).  Therefore, the matrix 
$B_3$ is similarly constrained to
\begin{equation}
B_3 =
\left( {\begin{array}{cc}
aI & bI \\
cI & dI \\
\end{array}} \right)\ .
\label{IIIeq14}
\end{equation}

For rotations only, the condition given by Eq. (\ref{IIIeq4}) produces 
four $2 \times 2$ conditions with the same results as those found for 
two-component spinors.  The matrices are constrained to 
\begin{equation}
B_{2j} =
\left( {\begin{array}{cc}
e\sigma_j & f\sigma_j \\
g\sigma_j & h\sigma_j \\
\end{array}} \right)\ .
\label{IIIeq15}
\end{equation}

Applying boosts (without rotations), and the condition given by 
Eq. (\ref{IIIeq3}) leads to $q=0$ and $t=p$, so that
\begin{equation}
B_1 =
\left( {\begin{array}{cc}
pI & 0 \\
sI & pI \\
\end{array}} \right).
\label{IIIeq16}
\end{equation}

In the case of boosts again without rotations, the condition 
given by Eq. (\ref{IIIeq4}) leads to $e=-h=s$ and $f=0$, so that
\begin{equation}
B_{2j} =
\left( {\begin{array}{cc}
s\sigma_j & 0 \\
g\sigma_j & -s\sigma_j \\
\end{array}} \right).
\label{IIIeq17}
\end{equation}

Applying boosts without rotations in the condition given by Eq. 
(\ref{IIIeq5}) leads to $b=2ims$, $p=0$, $a=d$, and $g=0$, and 
the set of matrices become
\begin{equation}
B_1 =
\left( {\begin{array}{cc}
0 & 0 \\
sI & 0 \\
\end{array}} \right),
\label{IIIeq18}
\end{equation}
\begin{equation}
B_{2j} =
\left( {\begin{array}{cc}
s\sigma_j & 0 \\
0 & -s\sigma_j \\
\end{array}} \right)\ ,
\label{IIIeq19}
\end{equation}
and
\begin{equation}
B_3 =
\left( {\begin{array}{cc}
aI & 2imsI \\
cI & aI \\
\end{array}} \right)\ .
\label{IIIeq20}
\end{equation}

Applying rotations and boosts in the conditions leads to 
$a=0$ and $c=0$, so that 
\begin{equation}
B_1 =
\left( {\begin{array}{cc}
0 & 0 \\
sI & 0 \\
\end{array}} \right)\ ,
\label{IIIeq21}
\end{equation}
\begin{equation}
B_{2j} =
\left( {\begin{array}{cc}
s\sigma_j & 0 \\
0 & -s\sigma_j \\
\end{array}} \right)\ ,
\label{IIIeq22}
\end{equation}
and
\begin{equation}
B_3 =
\left( {\begin{array}{cc}
0 & 2imsI \\
0 & 0 \\
\end{array}} \right)\ .
\label{IIIeq23}
\end{equation}

The constant $s$ can now be factored out of the equation 
leaving the following first-order differential equation
\begin{equation}
\left[
\left( {\begin{array}{cc}
0 & 0 \\
I & 0 \\
\end{array}} \right)
 {{\partial} \over {\partial t}}+
\left( {\begin{array}{cc}
\sigma_j & 0 \\
0 & -\sigma_j \\
\end{array}} \right)
{{\partial} \over {\partial x_j}} +
\left( {\begin{array}{cc}
0 & 2imI \\
0 & 0 \\
\end{array}} \right)
\right] \psi(\vec x,t)= 0\ ,
\label{IIIeq24}
\end{equation}

\noindent
which is the Galilean invariant first-order dynamical equation for 
four-component spinors.  This concludes the proof of Proposition 2.  

In the literature$^8$, Eq. (\ref{IIIeq24}) is known as the 
L\'evy-Leblond equation.  It is important to point out here that 
our derivation of this equation presented in Proposition 2 is 
different than that originally used by L\'evy-Leblond (see Refs
13 and 14).  Actually, the obtained equation can be cast into 
several different but equivalent forms.  This can be done by 
similarity transformations, which corresponds to a change of 
basis.  The equation can also be transformed into other 
representations such as momentum representation$^{13}$.

\subsection{Higher-order L\'evy-Leblond equations}

We also considered higher-order L\'evy-Leblond equations, which 
were derived by raising the original L\'evy-Leblond equation to
higher powers.  The following proposition summarizes the obtained 
results.

\bigskip
\noindent
{\bf Proposition 3.}{\it The L\'evy-Leblond equation with the 
operator raised to $N$-th power
\begin{equation}
\left[
\left( {\begin{array}{cc}
0 & 0 \\
I & 0 \\
\end{array}} \right)
 {{\partial} \over {\partial t}}+
\left( {\begin{array}{cc}
\sigma_j & 0 \\
0 & -\sigma_j \\
\end{array}} \right)
{{\partial} \over {\partial x_j}} +
\left( {\begin{array}{cc}
0 & 2imI \\
0 & 0 \\
\end{array}} \right)
\right]^N \psi(\vec x,t) = 0
\label{IIIeq25}
\end{equation}

\noindent
is Galilean invariant.}

\bigskip
\noindent
{\bf Proof.} Let 
\begin{equation}
\mathcal{L}=
\left( {\begin{array}{cc}
0 & 0 \\
I & 0 \\
\end{array}} \right)
 {{\partial} \over {\partial t}}+
\left( {\begin{array}{cc}
\sigma_j & 0 \\
0 & -\sigma_j \\
\end{array}} \right)
{{\partial} \over {\partial x_j}} +
\left( {\begin{array}{cc}
0 & 2imI \\
0 & 0 \\
\end{array}} \right)
\label{IIIeq26}
\end{equation}

\noindent
be the L\'evy-Leblond operator.  It has already been proven that 
the first-order equation is invariant (see Proposition 2), therefore 
we have
\begin{equation}
G\mathcal{L} G^{-1}G\psi(\vec x, t) =G\mathcal{L} G^{-1}e^{i\phi}
\psi(\vec x, t) =e^{i\phi}\mathcal{L}\psi(\vec x, t) = 0\ .
\label{IIIeq27}
\end{equation}

\noindent
This process can be repeated for each power of $\mathcal{L}$ until 
the phase factor has been commuted fully to the left, and the result 
is
\begin{equation}
G\mathcal{L}^N G^{-1}T\psi(\vec x, t)  =G\mathcal{L}^{N-1} G^{-1}
e^{i\phi}\mathcal{L}\psi(\vec x, t)=e^{i\phi}\mathcal{L}^N
\psi(\vec x, t)=  0\ .
\label{IIIeq28}
\end{equation}

\noindent
This concludes the proof of Proposition 3.

In the special case of $N = 2$, Eq. (\ref{IIIeq25}) gives the 
the Schr\"{o}dinger equation$^{17}$.  Hence, we can formulate 
the following corollary. 

\bigskip
\noindent
{\bf Corollary.} {\it The L\'evy-Leblond equation with 
the operator raised to $N = 2$ power is equivalent to 
the Schr\"{o}dinger equation.}

\bigskip
The resulting Schr\"{o}dinger equation can be written as
\[
\mathcal{L}^2 \psi(\vec x,t)=
\left[
\left( {\begin{array}{cc}
0 & 0 \\
I & 0 \\
\end{array}} \right)
 {{\partial} \over {\partial t}}+
\left( {\begin{array}{cc}
\sigma_j & 0 \\
0 & -\sigma_j \\
\end{array}} \right)
{{\partial} \over {\partial x_j}} +
\left( {\begin{array}{cc}
0 & 2imI \\
0 & 0 \\
\end{array}} \right)
\right]^2 \psi(\vec x,t)
\]
\[
=
\left[
2im\left( {\begin{array}{cc}
I & 0 \\
0 & I \\
\end{array}} \right)
 {{\partial} \over {\partial t}}+
\left( {\begin{array}{cc}
\{\sigma_j, \sigma_k\} & 0 \\
0 & \{\sigma_j, \sigma_k\} \\
\end{array}} \right)
{{\partial} \over {\partial x_j}} 
{{\partial} \over {\partial x_k}}
\right] \psi(\vec x,t) 
\]
\[
=
\left[
2im\left( {\begin{array}{cc}
I & 0 \\
0 & I \\
\end{array}} \right)
 {{\partial} \over {\partial t}}+2\delta_{jk}
\left( {\begin{array}{cc}
I & 0 \\
0 & I \\
\end{array}} \right)
{{\partial} \over {\partial x_j}} 
{{\partial} \over {\partial x_k}}
\right] \psi(\vec x,t)
\]
\begin{equation}
=
\left[
imI {\partial_t}+I{\partial_j^2} 
\right] \psi(\vec x,t)=0\ ,
\label{IIIeq29}
\end{equation}

\noindent
which shows that each component of the four-component spinor state 
function $\psi(\vec x,t)$ obeys the Schr\"{o}dinger equation, and 
that the latter does not mix the spinor components.   

\subsection{Fundamental dynamical equations for four-component spinors}

We have already shown that the only Galilean invariant first-order 
dynamical equation for four-component spinor state functions is the 
L\'evy-Leblond equation (see Proposition 2).  Furthermore there is an 
entire class of higher-order L\'evy-Leblond equations, which result 
from taking the $N^{th}$ power of the L\'evy-Leblond operator, and
these equations are Galilean invariant (see Proposition 3).  An 
interesting result is that among this class of equations, there is 
the Schr\"odinger equation$^{17}$, which is obeyed by all spinor 
components, however, it does not mix them.  

We now present a formal proof that there are no other invariant 
dynamical equations for four-component spinor state functions in
Galilean space-time.  This is a new result and one of the most 
important of this paper, so we present it as the following 
proposition. 

\bigskip
\noindent
{\bf Proposition 4.}{\it If a four-component spinor state function 
$\psi$ transforms as $\psi(\vec x,t) = e^{i\phi(\vec x^{\prime},
t^{\prime})} \psi^{\prime}(\vec x^{\prime}, t^{\prime})$, where 
$\phi$ is the Schr\"odinger phase function given by Eq. (\ref{IIeq9}),
and $(\vec x, t)$ and $(\vec x^{\prime}, t^{\prime})$ are two inertial 
frames of reference, then the L\'evy-Leblond and Schr\"odinger equations
are the only fundamental (Galilean invariant) dynamical equations.
}

\bigskip
\noindent
{\bf Proof:} 
Let us first consider equations among the L\'evy-Leblond class.  The $N$th 
order equation is obtained by raising the L\'evy-Leblond operator to the 
$N$th power.  The even and odd powers can be examined separately such that
\begin{equation}
\mathcal{L}^N=\left \{ \begin{array}{c}
\mathcal{L}^{2M}\ \ \ \  N even \\
\mathcal{L}^{2Q+1}\ \ \ \  N odd\\
\end{array} \right.
\label{IIIeq30}
\end{equation}

\noindent
where $M=N/2 \geq 1$ and $Q=(N-1)/2 \geq 1$.  Using the result given by 
Eq. (\ref{IIIeq29}) for $N=2$ and expanding the binomial to power $M$ 
produces 
\begin{equation}
\mathcal{L}^{2M}=\left[2imI\partial_t+2I\partial_j^2 \right]^M
=\sum^{p+q=M}_{p, q} {M! \over p!q!} (2im\partial_t)^p(2\partial_j)^q\ .
\label{31}
\end{equation}

Since mixed partial derivatives are produced for every term where $p>0$ 
and $q>0$, and since such mixed derivatives are not allowed in fundamental
equations (see Sec. 1), then there are no fundamental equations for $M>1$.  
When $N$ is odd, then
\begin{equation}
\mathcal{L}^{2Q+1} = \sum^{p+q=Q}_{p,q}{Q! \over p!q!}(2im
\partial_t)^p(2\partial_j)^q\ .
\label{32}
\end{equation}

\noindent
In this case, the mixed partial differentials result for every $Q>0$.  
There is no fortuitous canceling of terms as it was found for $N=2$,
which gave the Schr\"odinger equation, so there are no other fundamental 
equations.  

Knowing that there are no other fundamental equations among the L\'evy-Leblond 
class does not rule out other $2$nd and higher-order fundamental equations.  
To prove that there are no other fundamental equations for four-component 
spinor state functions, we must consider the most arbitrary $N^{th}$ order 
differential equation
\begin{equation}
\left[\sum_{a,b}^{a+b \leq N}D_{abj}\partial_t^a \partial_j^b \right]
\psi(\vec x, t)=0\ ,
\label{IIIeq33}
\end{equation}

\noindent
where $D_{abj}$ are $4 \times 4$ constant matrices and there is an implied 
summation over the index $j = 1, 2, 3$.  Eliminating mixed partial differential 
terms, Eq. (\ref{IIIeq33}) becomes
\begin{equation}
\left[\sum_{a}^{0<a \leq N}D_{a0}\partial_t^a + \sum_{b}^{0<b \leq N}
D_{0bj} \partial_j^b + D_{00} \right]\psi(\vec x, t)=0\ .
\label{IIIeq34}
\end{equation}

The Galilean transformation rule (see Eqs \ref{IIeq4} and \ref{IIeq5}) can be written
\[G{\partial_t}^p{{\partial }_i}^q{G}^{-1}e^{i\varphi ({\vec x},t)}\ 
\psi \left({\vec x},t\right)\] 
\begin{equation}
=e^{i\varphi ({\vec x},t)}{\left[k_1+\partial_t+k_{2i}\partial_i\right]}^p
{\left[k_{3i}+ k_{4ji}\partial_j\right]}^q\ \psi \left({\vec x},t\right)\ ,
\label{IIIeq35}
\end{equation} 

\noindent
where the introduced constants are $k_1=-i mv^2/2$, $k_{2i}=R_{ji}v_j$, 
${k_{2i}}^2={\left(R_{ji}v_j\right)}^2=v^2$, $k_{3i}=-im{R_{ji}v}_kR_{jk}$,
${k_{3i}}^2={\left(-im{R_{ji}v}_kR_{jk}\right)}^2=-m^2v^2$, and $k_{4ji}= 
R_{ji}$.  According to this rule Eq. (\ref{IIIeq34}) transforms into
\[
e^{i\varphi ({\vec x},t)} \left[\sum_{a}^{0<a \leq N}D'_{a0}\left(k_1+
\partial_t+k_{2i}\partial_i\right)^a \right.
\]
\begin{equation}
\left.+ \sum_{b}^{0<b \leq N}D'_{0bj}\left(k_{3j}+k_{4ij}\partial_i\right)^b 
+ D'_{00} \right]\psi(\vec x, t)=0\ .
\label{IIIeq36}
\end{equation}

The trinomial and binomials can be expanded by their respective powers
\[      
e^{i\varphi ({\vec x},t)} \left[\sum_{a}^{0<a \leq N} D'_{a0} \left( 
\sum_{p,q,r}^{p+q+r=a} P\left(k_1^p,1^q,k_{2i}^r \right) \partial_t^q 
\partial_i^r \right) \right.
\]
\begin{equation}
\left.+ \sum_{b}^{0<b \leq N}D'_{0bj}\left(\sum_{u,v}^{u+v=b} P\left( 
k_{3j}^{u},k_{4ij}^{v}\right) \partial_i^v \right) + D'_{00} \right]
\psi(\vec x, t)=0\ ,
\label{IIIeq37}
\end{equation}

\noindent
where the function $P()$ produces the permutation of its arguments.  To 
be Galilean invariant this equation must be equal to the untransformed 
equation (see Eq. \ref{IIIeq34}) and terms with mixed partial differentials 
must vanish.  The condition that must be met for the sum of mixed partial 
differential terms of like powers $e$ and $f$ to vanish is
\begin{equation}
\sum_{a}^{0<a \leq N} D'_{a0} \left( \sum_{p}^{p+e+f=a} P\left(k_1^p,1^e,
k_{2i}^f \right) \right)=0\ ,
\label{IIIeq38}
\end{equation}

\noindent
with $e>0$ and $f>0$.

The permuted terms do not vanish so the matrices must sum together to equal 
zero.  Hence, the remaining conditions for invariance are
\begin{equation}
D_{e0}=\sum_{a}^{0<a\leq N}D'_{a0}  \left( \sum_{p}^{p+e=a} P\left(k_1^p,1^e,
k_{2i}^0 \right) \right)\ ,
\label{IIIeq39}
\end{equation}
\[
D_{0fi}=\sum_{a}^{0<a\leq N}D'_{a0}  \left( \sum_{p}^{p+f=a} P\left(k_1^p,
1^0,k_{2i}^f \right) \right)
\]
\begin{equation}
\ \ \ \ \ \ \ \ + \sum_{b}^{0<b \leq N}D'_{0bj}\left(\sum_{u}^{u+f=b} 
P\left(k_{3j}^{u},k_{4ij}^{f}\right) \right)\ ,
\label{IIIeq40}
\end{equation}
and
\[
D_{00}=\sum_{a}^{0<a \leq N} D'_{a0} \left( P\left(k_1^a,1^0,k_{2i}^0 
\right) \right)
\]
\begin{equation}
\ \ \ \ \ \ \ \ + \sum_{b}^{0<b \leq N}D'_{0bj}\left( P\left(k_{3j}^{b},
k_{4ij}^{0}\right) \right) + D'_{00}\ .
\label{IIIeq41}
\end{equation}

Setting $f=0$ in Eq. (\ref{IIIeq38}) and combining with Eq. (\ref{IIIeq39}) 
proves that $D_{a0}=0$.  So there are no other Galilean invariant dynamical 
equations that are free of mixed partial differentials for 4-component 
spinors.  This concludes the proof of Proposition 4. 

\subsection{Discussion}

There are several important results of this paper.  First, we demonstrated 
that there are no fundamental dynamical equations for two-component spinor 
wave functions in Galilean space-time.  Second, we derived the L\'evy-Leblond 
equation for a four-component spinor wave function by using a different method 
than the original approach used by L\'evy-Leblond$^{13,14}$ and Fushchich and 
Nikitin$^8$.  Third, we presented a formal proof that the L\'evy-Leblond equation 
is the only first-order fundamental dynamical equation for four-component spinor 
wave functions in Galilean space-time.  Finally, we also proved that the 
Schr\"odinger equation is the only second-order fundamental dynamical equation 
in Galilean space-time, and that there are no other higher-order fundamental 
equations for two and four-component spinors.  It must be noted that all our 
results were obtained based on the Schr\"odinger phase factor, and that these 
results have far reaching physical consequences that will be now discussed.  

Quantum mechanics textbooks$^{20,30}$ frequently present the Pauli-Schr\"odinger 
(PS) equation for two-component spinors as a way of introducing the concept of 
spin.  Rotational invariance is expected of the equation and rotations are 
applied to the spinor state functions.  However it is not possible to do the 
same for Galilei boosts.  Consequently, in non-relativistic quantum mechanics 
there is no way to relate the spinor state function of one observer to that of 
another observer moving with a constant relative velocity.  Furthermore even the 
PS equation with its identity matrix coefficients cannot be shown to be Galilean 
invariant for two-component spinors because there are no $2 \times 2$ Galilei 
boost matrices (see Proposition 1).  We further discuss this problem in Sec. 4.

Now, our results demonstrate that the Schr\"odinger equation for four-component 
spinor state functions is Galilean invariant (see Proposition 3 and Corollary 
that follows it).  As a result, each spinor component must obey this equation.  
On the other hand, the equation does not mix the spinor component, which means 
that it does not lead to any new results.  Hence, our results allow us to conclude
that the L\'evy-Leblond equation for four-component spinor state functions is 
the only fundamental dynamical equation that correctly describes elementary 
particles with spin $1/2$ in Galilean space-time.   

\section{Pauli-Schr\"odinger equation}

The fact that the Dirac, Pauli-Schr\"odinger (PS) and Schr\"odinger equation 
are intimately related is well-known$^{31-33}$.  The PS equation is an 
approximation to the Dirac equation for small electron velocities and the 
Schr\"odinger equation can be obtained from the PS equation by neglecting 
magnetic interaction of the spin.  Since the PS equation describes evolution 
of a two-component spinor state function in Galilean space-time, it is used 
to introduce spin in non-relativistic quantum mechanics.  The equation can 
be formally derived from the L\'evy-Leblond equation$^{13}$.  We now discuss 
the relationship between the obtained results and the PS equation.

Let us introduce the four-potential $(V, A_j)$ of the electromagnetic field, 
where $V$ and $A_j$ are the scalar and vector potentials, and make the 
following substitutions:
\begin{equation}
i \partial_t \rightarrow i \partial_t-V(\vec x,t)\ ,
\label{IVeq1}
\end{equation}
and
\begin{equation}
-i\partial_j \rightarrow -i \partial_j - A_j (\vec x,t)\ . 
\label{IVeq2}
\end{equation}

\noindent
Maxwell's equations break Galilean invariance but may be cast into a Galilean 
invariant form by the elimination of Maxwell's term in the non-relativistic 
limit ($c \rightarrow \infty$).  Performing the above substitution in the 
L\'evy-Leblond equation produces a pair of coupled equations of the form
\begin{equation}
\sigma_j(i \partial_j+A_j) \phi - 2m \chi = 0\ ,
\label{IVeq3}
\end{equation}
and
\begin{equation}
(i \partial_t -v) \phi + \sigma_j(i \partial_j + A_j) \chi = 0\ .
\label{IVeq4}
\end{equation}

Then the Pauli-Schr\"odinger equation is obtained by eliminating $\chi$ 
from the above pair of equations, and we have
\begin{equation}                                                                                  i \partial_t \phi = V \phi + {1 \over 2m} \left[ (i \partial_j +A_j)^2 
+ i \sigma \cdot (i \partial_j+A_j)\times  (i \partial_j+A_j)\right]\phi\ .
\label{IVeq5}
\end{equation}

The PS equation is a second order differential equation governing the space 
and time evolution of a two-component spinor.  As a two-component spinor 
equation it cannot be proven to be Galilean invariant as demonstrated by 
Proposition 1.  This means that the PS equation is not fundamental in 
Galilean space-time.  It is interesting that the 2-component PS equation 
can be derived from the four-component L\'evy-Leblond equation, which is 
fundamental.  Thus the validity of the two-component PS equation follows 
from the existence of the fundamental (Galilean invariant) L\'evy-Leblond 
equation.

It has been suggested that the PS equation is covariant in the low velocity 
limit$^{33}$.  However that proof uses the four-component boost matrix to 
determine the effect of a boost on the separable two-components of the spinor.  
Normally the four-component boost matrix mixes the components of the spinor but 
in the limit of low velocity the four-component spinor boost matrix becomes an 
identity matrix and no mixing occurs.  As such it effectively applies no boost 
at all.  Consequently, this approach does not show Galilei boost invariance 
in a low velocity limit.

Based on the above results, some confusion may arise from the unusual fact that 
the four-component Schr\"odinger equation is fundamental while the two-component 
PS equation is not.  Since the matrices of the PS equation are diagonal there is 
no mixing of the pair of two-component spinors in the four-component equation.  
This makes it possible to separate the equations into two two-component PS 
equations$^{30}$, which indicates that the validity of the (non-fundamental) 
two-component PS equation is a consequence of the fundamental four-component 
Schr\"odinger equation (see Corollary following Proposition 3).

The above discussion of the PS equation clearly shows that the L\'evy-Leblond 
and Schr\"odinger equations for four-component spinor wave functions are the 
only fundamental dynamical equation in Galilean space-time with the Schr\"odinger 
phase factor (see Proposition 4).  This is an important result as it shows that 
only the L\'evy-Leblond and Schr\"odinger equations are available to formulate 
quantum field theories of elementary particles described by four-component 
spinor state functions in Galilean space-time.  

\section{Conclusions}

A search for fundamental dynamical equations for two and four-component spinor 
state functions was conducted in Galilean space-time represented by the extended 
Galilei group.   For a dynamical equation to be considered fundamental, it was 
required that the equation was invariant under the symmetry operators of the 
group of the Galilei metric, that the state functions transformed like the 
irreducible representations of the group of the metric, and that the equation 
did not have mixed time and space partial derivatives.  

The main results obtained are: (i) there are no fundamental dynamical 
equations for two-component spinor wave functions in Galilean space-time;  
(ii) the L\'evy-Leblond equation for a four-component spinor wave function 
can be derived by using a different method than the one originally used by 
L\'evy-Leblond; (iii) a formal proof that the L\'evy-Leblond equation is the 
only first-order fundamental dynamical equation for four-component spinor wave 
functions in Galilean space-time; (iv) the Schr\"odinger equation is the only 
second-order fundamental dynamical equation in Galilean space-time; (v) there 
are no other higher-order fundamental equations for two and four-component 
spinors; and (vi) the Pauli-Schr\"odinger equation for two-component spinor 
wave functions is not fundamental and that this is a consequence of the 
fundamental Schr\"odinger equation for four-component spinor state functions. 
Our conclusions (iii) and (iv) are similar to those reached by Fushchich and 
Nikitin$^8$, and more recently by de Montigny, Niederle and Nikitin$^{24}$ 
and by Niederle and Nikitin$^{25}$, albeit by a different analysis.

All our results presented here were obtained with the Schr\"odinger phase 
function$^{20}$, however, other phase functions are also possible.  It 
will be shown in the next paper of this series that such phase functions  
lead to new fundamental dynamical equations that have important physical
implications. 

\bigskip\noindent
{\bf ACKNOWLEDGMENTS}
We thank Alex Weiss and Chris Jackson for discussions and their 
comments on the manuscript. Z.E.M. acknowledges the support of 
this work by the Alexander von Humboldt Foundation.  
%

\end{document}